%Paper: hep-th/9505105
%From: vafa@string.harvard.edu (Cumrun Vafa)
%Date: Wed, 17 May 1995 09:45:07 -0400
%Date (revised): Wed, 24 May 1995 15:38:11 -0400
%Date (revised): Tue, 1 Aug 95 11:24:27 -0400

\input harvmac.tex
\noblackbox
\lref\KLM{A. Klemm, W. Lerche, and P. Mayr, ``K3-Fibrations and
Heterotic-Type II String Duality,'' hep-th/9506112.}
\lref\VaW{C. Vafa and E. Witten, ``Dual String Pairs with N=1 and
N=2 Supersymmetry in Four Dimensions,'' hep-th/9507050.}
\lref\Yautwo{S. Hosono, A. Klemm, S. Theisen, and S.T. Yau,
``Mirror Symmetry, Mirror Map, and Applications to Complete Intersection
Calabi-Yau Spaces,'' {\it Nucl. Phys.} {\bf B433} (1995) 501, hep-th/9406055.}
\lref\level{C. Vafa, ``Modular Invariance and Discrete Torsion on Orbifolds,''
{\it Nucl. Phys.} {\bf B273} (1986) 592; G. Athanasiu, J. Atick,
M. Dine, and W. Fischler, ``Remarks on Wilson Lines, Modular Invariance,
and Possible String Relics in Calabi-Yau Compactifications,''
{\it Phys. Lett.} {\bf 214B} (1988) 55.}
\lref\Schimm{R. Schimmrigk, ``Heterotic RG Flow Fixed Points
with Nondiagonal Affine Invariants,'' {\it Phys. Lett.} {\bf B229}
(1989) 227.}
\lref\HYPMirror{D. Morrison, ``Picard-Fuchs Equations and Mirror Maps for
Hypersurfaces,'' in {\it Essays on Mirror Manifolds}, ed. S.T. Yau,
International Press, 1992, alg-geom/9202026; A. Font, ``Periods and
Duality Symmetries in Calabi-Yau Compactifications,'' {\it Nucl. Phys.}
{\bf B391} (1993) 358, hep-th/9203084; A. Klemm and S. Theisen,
``Considerations of One Modulus Calabi-Yau Compactification: Picard
Fuchs Equations, Kahler Potentials, and Mirror Maps,'' {\it Nucl.
Phys.} {\bf B389} (1993) 153, hep-th/9205041.}
\lref\oldS{A. Font, L. Ibanez, D. Lust, and F. Quevedo, ``Strong-Weak
Coupling Duality and Nonperturbative Effects in String Theory,''
{\it Phys. Lett.} {\bf B249} (1990) 35; J. Schwarz, ``Does String
Theory have a Duality Symmetry Relating Weak and Strong Coupling,''
hep-th/9307121, and references therein.}
\lref\confo{ P. Candelas, A.M. Dale, C.A. L\"utken, and R. Schimmrigk,
``Complete Intersection Calabi--Yau Manifolds'', Nucl. Phys. B298 (1988)
493--525;  P.S. Green and T.  H\"ubsch,
``Possible Phase Transitions Among Calabi--Yau Compactifications'',
Phys. Rev. Lett. 61 (1988) 1163--1166;
``Connecting Moduli Spaces of Calabi--Yau Threefolds'', Comm. Math.
Phys. 119 (1988) 431--441;
P. Candelas, P.S. Green, and T.
H\"ubsch,
``Finite Distance Between Distinct Calabi--Yau Manifolds'',
Phys. Rev. Lett. 62 (1989) 1956--1959;
``Rolling Among Calabi--Yau Vacua'', Nucl. Phys.
 B330 (1990) 49--102;
 P. Candelas and X.C. de la Ossa,
``Comments on Conifolds'', Nucl. Phys. B342 (1990) 246--268.}
\lref\Candnew{P. Green, T. Hubsch, and C. Lutken, ``All the Hodge Numbers
for all Calabi-Yau Complete Intersections,'' {\it Class. Quantum Grav.}
{\bf 6} (1989) 105.}
\lref\Narain{K. Narain, ``New Heterotic String Theories in Uncompactified
Dimensions $<$ 10,'' {\it Phys. Lett.} {\bf 169B} (1986) 41.}
\lref\Gepner{D. Gepner, ``Exactly Solvable String Compactifications on
Manifolds of SU(N) Holonomy,'' {\it Phys. Lett.} {\bf 199B} (1987) 380.}
\lref\Schwarz{M. Green, J. Schwarz, and P. West, ``Anomaly-Free Chiral
Theories in Six Dimensions,'' {\it Nucl. Phys.} {\bf B254} (1985) 327.}
\lref\kap{V. Kaplunovsky and J. Louis, ``On Gauge Couplings in
String Theory,'' hep-th/9502077.}
\lref\SWTwo{N. Seiberg and E. Witten, ``Monopoles, Duality, and
Chiral Symmetry Breaking in N=2 Supersymmetric QCD,'' {\it Nucl. Phys.}
 { \bf B431} (1994) 484, hep-th/9408099.}
\lref\othern{A. Klemm, W. Lerche, S. Yankielowicz, and S. Theisen, ``Simple
Singularities and N=2 Supersymmetric Yang-Mills Theory,'' {\it Phys. Lett.}
{\bf B344} (1995) 169, hep-th/9411048.}
\lref\otherntwo{
P. Argyres and A. Faraggi, ``The Vacuum Structure
and Spectrum of N=2 Supersymmetric SU(N) Gauge Theory,'' hep-th/9411057.}
\lref\VW{C. Vafa and E. Witten, ``A One Loop Test of String Duality,''
hep-th/9505053.}
\lref\largen{M. Douglas and S. Shenker, ``Dynamics of SU(N) Supersymmetric
Gauge Theory,'' hep-th/9503163.}
\lref\three{P. Argyres and M. Douglas, ``New Phenomena in SU(3)
Supersymmetric Gauge Theory,'' hep-th/9505062.}
\lref\asymm{K. Narain, M. Sarmadi, and C. Vafa, ``Asymmetric Orbifolds,''
{\it Nucl. Phys.} {\bf B288} (1987) 551.}
\lref\Vafatest{C. Vafa, ``A Stringy Test of the Fate of the Conifold,''
hep-th/9505023.}
\lref\ferrara{A. Ceresole, R. D'Auria, S. Ferrara, and A. Van Proeyen,
``On Electromagnetic Duality in Locally Supersymmetric N=2 Yang-Mills
Theory,'' hep-th/9412200.}
\lref\ceresole{A. Ceresole, R. D'Auria, and S. Ferrara, ``On the
Geometry of the Moduli space of Vacua in N=2 Supersymmetric Yang-Mills
Theory,'' {\it Phys. Lett.} {\bf B339} (1994) 71, hep-th/9408036.}
\lref\louis{B. de Wit, V. Kaplunovsky, J. Louis, and D. Lust,
``Perturbative Couplings of Vector Multiplets in N=2 Heterotic String Vacua,''
hep-th/9504006.}
\lref\Witt{E. Witten, ``New Issues in Manifolds of SU(3) Holonomy,''
{\it Nucl. Phys.} {\bf B268} (1986) 79.}
\lref\oldstrom{A. Strominger, ``Superstrings with Torsion,'' {\it Nucl.
Phys.} {\bf B274} (1986) 253.}
\lref\Candelas{P. Candelas, X. De la Ossa, A. Font, S. Katz, and D. Morrison,
``Mirror Symmetry for Two Parameter Models - I,'' {\it Nucl. Phys.}
{\bf B416} (1994) 481, hep-th/9308083.}
\lref\Harvey{J. Harvey and A. Strominger, ``The Heterotic String is a
Soliton,'' hep-th/9504047.}
\lref\Sen{A. Sen, ``String-String Duality Conjecture in Six Dimensions
and Charged Solitonic Strings,'' hep-th/9504027.}
\lref\Hull{C. Hull and P. Townsend, ``Unity of Superstring Dualities,''
hep-th/9410167.}
\lref\Witten{E. Witten, ``String Theory Dynamics in Various Dimensions,''
hep-th/9503124.}
\lref\strom{A. Strominger, ``Massless Black Holes and Conifolds in
String Theory,'' hep-th/9504090.}
\lref\stromtwo{B. Greene, D. Morrison, and A. Strominger, ``Black Hole
Condensation and the Unification of String Vacua,'' hep-th/9504145.}
\lref\Antoniadis{I. Antoniadis, S. Ferrara, E. Gava, K.S. Narain, and
T.R. Taylor, ``Perturbative Prepotential and Monodromies in N=2
Heterotic String,'' hep-th/9504034.}
\lref\Yau{S. Hosono, A. Klemm, S. Theisen, and S.T. Yau, ``Mirror Symmetry,
Mirror Map and Applications to Calabi-Yau Hypersurfaces,''
Comm. Math. Phys. {\bf 167} (1995) 301, hep-th/9308122.}
\lref\Seiberg{N. Seiberg and E. Witten, ``Electric-magnetic Duality,
Monopole Condensation, and Confinement in N=2 Supersymmetric Yang-Mills
Theory,'' {\it Nucl. Phys.} {\bf B426} (1994) 19, hep-th/9407087.}
\lref\knownCY{P. Candelas, M. Lynker, and R. Schimmrigk,
``Calabi-Yau Manifolds in Weighted P4,'' {\it Nucl. Phys.} {\bf B341}
(1990) 383; A. Klemm and R. Schimmrigk, ``Landau-Ginzburg String Vacua,''
{\it Nucl. Phys.} {\bf B411} (1994) 559, hep-th/9204060; M. Kreuzer and
H. Skarke, ``No Mirror Symmetry in Landau-Ginzburg Spectra!,''
{\it Nucl. Phys.} {\bf B388} (1992) 113, hep-th/9205004.}
\lref\DK{J. Distler and S. Kachru, ``Duality of (0,2) String Vacua,''
hep-th/9501111, to appear in Nucl. Phys. {\bf B}.}
\lref\senrev{A. Sen, ``Strong-Weak Coupling Duality in Four-Dimensional
String Theory,'' {\it Int. J. Mod. Phys.} {\bf A9} (1994) 3707,
hep-th/9402002.}
\lref\Walton{M. Walton, ``The Heterotic String on the Simplest Calabi-Yau
Manifold and its Orbifold Limits,'' {\it Phys. Rev.} {\bf D37} (1987)
377.}
\lref\Katz{P. Candelas, A. Font, S. Katz, and D. Morrison, ``Mirror Symmetry
for Two Parameters Models, II,'' {\it Nucl. Phys.} {\bf B429} (1994) 626,
hep-th/9403187.}
\Title{\vbox{\hbox{HUTP-95/A016}\hbox{\tt hep-th/9505105}}}
{Exact Results for N=2
 Compactifications of Heterotic Strings}
\bigskip
\centerline{Shamit Kachru\footnote{$^\dagger$}
{Junior Fellow, Harvard Society of Fellows} and Cumrun Vafa}
\bigskip\centerline{\it Lyman Laboratory of Physics}
\centerline{\it Harvard University}\centerline{\it Cambridge, MA 02138}

\vskip .3in

We search for $N=2$, $d=4$ theories which can be realized
both as heterotic string
compactifications on $K_{3}\times T^{2}$ and
as type II string compactifications
on Calabi-Yau threefolds.   In such cases, the exact
non-perturbative superpotential
of one string theory is given in
terms of tree level computations in the other string theory.
In particular we find concrete examples which provide the
stringy realization of the
results of Seiberg and Witten on N=2 Yang-Mills theory,
corrected by gravitational/stringy effects.
We also discuss some examples which shed light on how the moduli spaces
of different N=2 heterotic vacua are connected.

\Date{\it {May 1995}} %replace this line by \draft  for preliminary
%%versions

%\draft

\newsec{Introduction}

Recently, there
has been tremendous progress in our understanding of N=2 supersymmetric
field theories \refs{\Seiberg,\SWTwo,\othern,\otherntwo,\largen,\three}
and N=2 supersymmetric type II string
compactifications \strom\stromtwo\Vafatest.
At the same time, our knowledge of strong-coupling
phenomena in string theory has been enriched by exciting work on S-duality
and six-dimensional string-string duality
\refs{\oldS,\senrev,\Hull,\Witten,\Sen,\Harvey,\VW}.

In this paper, we explore N=2 supersymmetric vacua of the heterotic
string in four-dimensions.  We find examples which suggest that many
naively distinct heterotic N=2 moduli spaces are in fact connected, in a way
which is very analogous to the way the
different type II Calabi-Yau compactifications
are connected \stromtwo\confo.  Moreover as we will argue
some $N=2$ vacua have dual realizations as both type II and
heterotic compactifications.
This has dramatic
implications, as has been suggested by Strominger.

 One of the key ideas leading
to the resolution of the conifold singularity is
the postulate by Strominger that
the absence of neutral perturbative couplings between
vector multiplets and hypermultiplets survives
nonperturbative string effects.  Since the dilaton
is part of a hypermultiplet in type II compactifications
on Calabi-Yau threefolds, it follows that the tree
level prepotential for vector multiplets is exact\foot{In general,
different superpotential terms receive contributions at only
a specific order in the genus expansion, depending
on their modular weight. The moduli space
geometry is fixed at genus 0.},
while that of the hypermultiplets might get corrected.
In the context of $N=2$ compactifications of
heterotic strings, since the dilaton now sits
in a vector multiplet, it follows from Strominger's
postulate that the moduli space of hypermultiplets is exact
at the tree level
while the vector multiplets can receive quantum corrections.
This also implies that if we find an $N=2$ string compactification
which has realizations both as a type II and as a heterotic string
compactification, then the exact prepotential for the vector multiplets
can
be computed using the type II realization at tree level, and the exact
hypermultiplet superpotential can be computed using the heterotic
realization at tree level.  In particular
for our examples with dual realizations we compute
the non-perturbative corrections to the prepotential for the
vector multiplets on the heterotic side, using the known
prepotentials
of Calabi-Yau threefolds (thus realizing
the speculation by many physicists that there should be a connection
between quantum moduli spaces of heterotic strings and special
geometry of Calabi-Yau threefolds \ceresole\othern\ferrara).
The examples we consider
include models which have $SU(2)$ as well as $SU(3)$ enhanced
gauge symmetry points at the heterotic string tree level.
This in particular gives
us a string realization, including
gravitational corrections, of \Seiberg .

It would be desirable to have more examples of this $N=2$
heterotic/type II duality.
In particular the string-string duality in 6 dimensions
relating type IIA compactifications on $K_3$ to toroidal
compactifications of heterotic strings may provide
a hint of how dual pairs can be constructed
more systematically \Witten , which seems to lead to other 4 dimensional
$N=2$ type II/heterotic dual examples
\ref\group{ S. Ferrara, J. Harvey, A. Strominger and C. Vafa,
to appear.}.

In \S2 we briefly review the construction of general
(0,4) heterotic string compactifications on
$K_3 \times T^{2}$ (in \Schwarz\ similar techniques were used
to construct anomaly free six dimensional chiral gauge theories).
We then discuss several examples in \S3, and point out that ``stringy''
enhanced gauge symmetry points may provide a way of connecting many or all
N=2 supersymmetric heterotic vacua (including
those on asymmetric orbifolds \asymm).
In \S4 we come to the main focus
of this paper.  We
construct heterotic theories for which we can propose a specific candidate
dual Type IIA string compactification on a Calabi-Yau manifold, and in some
cases we give strong evidence for such a duality.
We give our conclusions in \S5.

The importance and implications of duality between $N=2$
compactifications of type II and heterotic strings, as well
as the dual interpretation of the conifold singularities as Seiberg-Witten
monopole points, has been independently noted by Ferrara, Harvey and
Strominger.

\newsec{Models on $K_{3}\times T^{2}$}

There are several different approaches one might take to constructing
N=2 heterotic string vacua.  The most straightforward is perhaps to
compactify the $E_{8}\times E_{8}$ or $SO(32)$ heterotic string on
$K_{3}\times T^{2}$.  We will first discuss this class of compactifications,
and then in \S3 will discuss modifications of such vacua
based on asymmetric orbifolds \asymm.

The most familiar way of compactifying the heterotic string on $K_{3}\times
T^{2}$ is simply to use the ``standard embedding,'' equating the spin
connection of the manifold with the gauge connection.  For the
$E_{8}\times E_{8}$ string this yields a theory (at
generic points in the moduli space of the torus) with $E_{7}\times E_{8}\times
U(1)^{4}$
gauge group, 10 hypermultiplets in the $\bf {56}$ of $E_{7}$, and
65 gauge neutral moduli hypermultiplets -- 20
moduli of the $K_3$ surface and 45 moduli
of the gauge bundle.

Of course, the adjoint scalar fields sitting in the vectors
corresponding to the Cartan subalgebra of
the gauge group are also moduli which can be given nonzero VEVs.
At $\it generic$ points in the moduli space of
these scalars,
the theory is characterized
by 65 hypermultiplets and 19 vectors (18
coming from vector multiplets including
the multiplet of the dilaton,
one coming from the graviphoton).
In the following we will denote this by $(65,19)$ and in general will
describe the spectrum of a theory
which at generic points has
M hypermultiplets and N vector fields (including
graviphoton) by
$(M,N)$.  We note now that in the context of Type IIA strings, such
a theory would arise from compactification
on a Calabi-Yau manifold with hodge numbers $b_{11}=N-1, b_{21}=M-1.$
The additional hypermultiplet and vector necessary to obtain agreement
with the heterotic spectrum would then come from the dilaton and
the graviphoton, respectively.

It is useful to review at this point how one could derive this spectrum
just using index theory.  Let us start with an unbroken gauge group ${\cal G}$
in ten dimensions
and break it to a subgroup $G$ by giving
gauge fields on $K_3$ an expectation value in $H$ where $G\times H \subset
{\cal G}$ is
a maximal subgroup.

Compactify further on a torus to get a four dimensional N=2 theory.
That part of the matter spectrum
which arises from the higher dimensional gauge multiplet can be determined
as follows.
Decompose
\eqn\adjoint{adj ~{\cal G} = \sum_{i} (M_{i},R_{i})~}
where $M_{i}$ and $R_{i}$ are representations of $G$ and $H$ respectively.
Then it follows from the index theorem that generically
the number of left-handed
spinor multiplets transforming in the $M_{i}$ representation
of $G$ is given by
\eqn\index{N_{M_{i}} ~=~\rm {\int_{K_3} -
{1\over 2} tr_{R_{i}} F^{2} + {1\over 48}
dim R_{i} ~tr R^{2}~=~dim R_{i} - {1\over 2} \int_{K_3} c_{2}(V)~
index(R_{i}).}}
Here $V$ denotes the $H$ bundle parametrizing the VEV of the vacuum
gauge fields on $K_{3}$.
In addition, for compactifications on $K_{3}$ there is a universal
contribution to the spectrum of matter hypermultiplets coming from the
higher dimensional gravitational fields; this consists of 20 gauge singlet
hypermultiplet moduli.

For the case of the standard embedding, we have chosen $V$ to be
an $SU(2)$ bundle with $\int_{K_{3}} c_{2}(V) = 24$.
Going through the computations
above yields the expected 10 $\bf {56}$s and 45 extra moduli hypermultiplets
arising from the higher dimensional gauge fields.
But in general we are free \Witt\oldstrom\ to choose more general
stable, holomorphic $SU(N)$ bundles over $K_{3}$
in the process of compactification (this corresponds to $H=SU(N)$ in the
previous discussion), subject only to the constraints
\eqn\bundcons{c_{2}(V) ~=~c_{2}(TK_{3}), ~~c_{1}(V) ~=~0.}
More generally, we may wish to choose several different
factors $V_{a}$ in the vacuum gauge
bundle and e.g. embed them in different $E_{8}$ factors.
In this case the constraints are simply
\eqn\gencons{\sum c_{2}(V_{a}) ~=~c_{2}(TK_{3}),~~c_{1}(V_{a})~=~0.}
When computing the generic spectrum, one only needs to know the number
of gauge neutral moduli hypermultiplets and the rank of the gauge group.
The gravitational
contribution is a universal $20$, while the number of moduli
of an $SU(N)$ bundle with $\int_{K_{3}}c_{2}(V) = A$ is
$AN + 1 - N^{2}$.
  Using these formulas (and the knowledge that embedding an $SU(N)$ bundle
in the gauge group will reduce the rank by $N-1$)
it is easy
to find the generic spectra of models of this type.

\newsec{Examples and Observations about their Moduli Spaces}

Let us discuss some of the different theories we may obtain in this way.
Embedding a single $SU(N)$ factor with $\int_{K_3} c_{2}(V) = 24$
in one $E_{8}$
of the heterotic string breaks this $E_{8}$ to
$E_{7}$, $E_{6}$, $SO(10)$, or $SU(5)$ for $N=2,3,4,5$ and results
in theories with
the following generic
spectra:
\eqn\specone{N=2:~(65,19)
{}~~~N=3:~(84,18) ~~~N=4:~(101,17) ~~~N=5:~(116,16)~.}
One can also compute the spectrum of charged fields in these models at the
points where nonabelian gauge symmetry is classically
restored, using the techniques outlined
in \S2.

The different models
listed in \specone\ are
of course not unrelated.  Starting with the $(65,19)$ model,
one knows that classically there is a point where an $E_{7}$ gauge symmetry
is restored and that the spectrum there includes 10 ${\bf 56}$s of $E_{7}$.
Under the
maximal $E_{6}\times U(1)$ subgroup of
$E_{7}$, $\bf {56} = {\bf {27}} + \bf {\overline {27}}
+ {\bf 1} + {\bf 1}$ with the $E_{6}$ singlets charged under the $U(1)$.
Moving to the codimension one locus in the moduli space of the Cartan vectors
where the scalar in that $U(1)$ vector multiplet
has vanishing VEV, one also therefore finds
an extra 20 massless hypermultiplets, charged under that $U(1)$.
One can now Higgs the U(1), leaving 19 extra gauge singlet fields --
in other words, one is now on a branch of moduli space with spectrum
$(84,18)$, the N=3 case of \specone.
One can similarly move from the N=3 case to the N=4 case and so forth.

It is very amusing to note the similarity between going from the N=4 to
N=5 case and the process described in \stromtwo\ moving from the
moduli space of a Calabi-Yau with hodge numbers $b_{11}=2, b_{21}=86$
to the moduli space of the quintic with $b_{11}=1, b_{21}=101$.
Note that these are the same numbers as we would have gotten
from $N=4$ and $N=5$ above except by an overall shift of 14 in both
$b_{11}$ and $b_{21}$.
The N=4 model has a special point in the moduli space of vectors where
there is an $SO(10)\times E_{8}\times U(1)^{4}$ gauge symmetry.
The charged spectrum there includes 16 $\bf {16}$s of $SO(10)$.
Under the $SU(5)\times U(1) \subset SO(10)$, one has
$\bf {16 = 10 + \bar 5 + 1}$ where the $\bf 1$ is charged under the
$U(1)$.  So on a locus of codimension one in the moduli space of the
vectors of the N=4 model, one finds 16 massless hypermultiplets of
fields charged under a single $U(1)$ gauge symmetry.  Higgsing this
gauge symmetry leads to 15 more neutral hypermultiplets and one less
vector multiplet.  This is precisely the ``mirror'' description of
the process described in \stromtwo!
The numbers of vector and
hypermultiplets are however shifted by 14, making it problematic to
conjecture a precise duality between this heterotic process and the
type II process described in \stromtwo.
However, if one shifts the numbers of vector and hypermultiplets by
14 in each of the N=2,3,4,5 cases listed in \specone\ one notices
that all of them yield numbers which would arise in type IIA
compactification on complete intersection Calabi-Yaus in products
of projective spaces \Candnew.  Since it is precisely this class of
manifolds which we know are connected by conifold transitions, it would
be interesting to see if one could somehow explain the shift of 14
and find a precise duality between these heterotic theories and some type II
examples.

We would like to find (0,4) heterotic compactifications for which
we $\it can$ find a dual type II compactification on a Calabi-Yau
threefold, and for which we can give a very stringent test of the duality.
While there are known Calabi-Yau manifolds with the requisite hodge numbers
to produce the numbers of vector and hypermultiplets of some of the
theories in \specone\ when used to compactify type II strings, these
examples are too complicated to
provide a good testing ground for such a duality conjecture.
We will come back to this point, and provide much better examples where
we can conjecture and give extremely strong evidence for such a duality.
But first, we find it worthwhile to discuss
the connectedness of the moduli space of (0,4) heterotic theories.

It is now strongly believed that type II compactifications on different
Calabi-Yau manifolds are connected smoothly \strom\stromtwo\ through conifold
transitions \confo.  In fact it has
been conjectured that
$\it all$ Calabi-Yau compactifications may be connected in this way.
In order to prove a similar statement
for moduli spaces of N=2 heterotic compactifications,
the classes of theories that one has to connect are even more disparate.

The simplest Calabi-Yau compactification which yields an N=2 heterotic
compactification is $K_{3}\times T^{2}$.  Different choices of gauge
bundles yield theories with different spectra, but we have seen
that often by moving to a special point in the vectors' moduli
space where charged hypermultiplets become massless, we can move to a new
partial Higgs phase and obtain a model with different numbers of
vectors and hypermultiplets.   Unlike the situation in \stromtwo, these
hypermultiplets arise in the perturbative spectrum of the heterotic string.

However, beyond the $K_{3}\times T^{2}$ compactifications,
there are many asymmetric
orbifold compactifications which yield N=2 heterotic theories in four
dimensions.  These are naively completely non-geometrical
(left and right movers live on different spaces!) and one
might despair of obtaining them as smooth deformations of $K_{3}\times T^{2}$
compactifications.

For example, one can easily write down, among many other
possibilities, orbifolds
with $(0,24)$ and $(4,20)$ which
one cannot obtain in the manner discussed
thus far by choosing stable bundles over $K_3$.
Consider for example a compactification on a Narain lattice
given by $\Gamma^{4,20}\oplus \Gamma^{2,2}$
where $\Gamma^{p,p+8k}$ are arbitrary self-dual
even lattices.
  If we choose $\Gamma^{4,20}$
to correspond to $SO(8)\times SO(40)$ weight lattices (with
difference in the root lattice \Narain) and just
consider the $Z_2$ reflection to act only on the $SO(8)$ part together
with a $v^2=0$ shift in $\Gamma^{2,2}$
we get a model with no hypermultiplets and with $22+2=24$
vector fields.  The $(4,20)$ model can be obtained
by considering $\Gamma^{4,4}\oplus \Gamma^{2,18}$ and modding
out by a reflection in $\Gamma^{4,4}$ accompanied by
a shift in $\Gamma^{2,18}$ (to make the
left-right level matching work).  This gives us $4$ hypermultiplets
from the moduli of $\Gamma^{4,4}$ and $2+18=20$ $U(1)$ gauge fields.
This latter model does correspond to a model
that can be realized geometrically as the holonomy action is left-right
symmetric.

The model with $(4,20)$ can also be obtained in another way once we recall
one of the most characteristic features
of the heterotic string -- special singular points in moduli space where
enhanced ``stringy'' gauge symmetries arise!
Consider for example the $K_3$
orbifold obtained as $T^{4}/Z_{2}$.  The spectrum
of this theory has been worked out in detail in \Walton.  At this orbifold
point an extra $SU(2)$ gauge symmetry appears.  Using the standard embedding
one finds hypermultiplets with the following
$E_{7}\times SU(2)$ charges:
\eqn\orbspec{8~(\bf{56}, \bf {1}), ~~{\rm 1}~(\bf{56},\bf{2}),~~{\rm 32}~
(\bf {1}, \bf {2}),~~
{\rm 4}~(\bf {1}, \bf {1})~.}
These include the familiar ten $\bf {56}$s and 65 moduli of the standard
embedding, but some of them are paired in SU(2) doublets.\foot{
The $SU(2)$ doublets pair some $K_3$ moduli with moduli of the
vacuum gauge bundle, showing that there is a duality analogous
to that of \DK\ for these $K_3$ theories.}
There are also three additional
Higgses for the SU(2) gauge symmetry.
If we break the SU(2) by Higgsing, we recover the $(65,19)$ theory discussed
above.  However, we can also give the scalar in the $U(1)\subset SU(2)$
vector multiplet a VEV, moving to the Coulomb phase of the enhanced
gauge symmetry.  This will give all of the SU(2) doublets masses, leaving
us with 20 vectors (from the Cartan piece of $E_{8}\times E_{7}\times SU(2)
\times U(1)^{4}$) and 4 hypermultiplets!
This reproduces the $(4,20)$ of the orbifold above.

Similarly, the $K_3$ moduli space at certain Gepner points develops an
extra rank 5 enhanced gauge symmetry \Gepner.
For example the $1^{6}$ Gepner
model has an extra $U(1)^{5}$ gauge symmetry\foot{
The same can be done using the $T^4/Z_2$ example above by choosing
the circles to correspond to $SU(2)$ symmetry points.}.  It is easy to check
that
all of the hypermultiplets are charged under one or more of these $U(1)$s,
so moving to a generic point in the moduli space of the vectors in this
theory yields a model with $(0,24)$, just like the asymmetric
orbifold example above.
Such examples make it natural to conjecture that most or all N=2 heterotic
compactifications are connected by such transitions from Higgs to Coulomb
phases, sometimes going through points with enhanced stringy gauge symmetries.

\newsec{Heterotic/Type II Duality}

\subsec{General Remarks}

As previously mentioned, some of the examples discussed so far do have
potential Calabi-Yau ``duals,'' but these manifolds are much too complicated
to allow for a really convincing check of any duality conjecture.
On the other hand, we have seen that by Higgsing using charged fields
we can reduce the rank of the unbroken gauge group.  If we reduce the rank
sufficiently, we obtain a model which would be dual to a type IIA
compactification on a Calabi-Yau with $b_{11}$ small.  Searching
for such Calabi-Yaus is easier because they are relatively rare.
Moreover for such examples, the exact structure of the moduli space of (1,1)
forms
has been determined using mirror symmetry\foot{
Alternatively, one could study type IIB strings
on the mirror, in which case the tree
level sigma model computes the exact structure of moduli space.},
and we could therefore give
stringent tests of any duality conjecture between the string tree level
moduli space of
(1,1) forms on such a Calabi-Yau and the exact quantum
moduli space of vector multiplets
in a given heterotic model.

We have already seen a hint of heterotic/type II duality in \S3.
There we saw special points in the moduli space of vectors where charged
hypermultiplets become massless; giving them VEVs Higgses the gauge symmetry
and moves us on to a new branch of the moduli space.  In the spirit of
\SWTwo, one would expect the ``quantum'' version of this story to change:
one would expect a ``magnetic Higgs phase'' in which charged solitons condense
to be responsible for the new branch of moduli space.  And the charged black
holes of \strom\stromtwo\ are amenable to exactly such an interpretation.
With these general remarks out of the way, let us move on to construct
some explicit examples of heterotic theories which are dual to type II
theories on Calabi-Yau threefolds.

In any N=2 heterotic string, there will be at least two vectors (one
vector multiplet) -- the graviphoton and the vector in the supermultiplet
of the dilaton.  If we want, classically, to have points with nonabelian
gauge symmetry, then we need at least one more vector, making it
desirable to study heterotic models with 3 vectors.  A type IIA string
compactified on a Calabi-Yau with $b_{11}=2$ would also give rise
to 3 vectors (including the graviphoton).  Similarly, a heterotic model in
which all of the gauge symmetry came from the $U(1)^{4}$ of the torus
(which is enhanced to nonabelian groups at special points) would be dual
to a Calabi-Yau with $b_{11}=3$, perhaps.
This makes the strategy clear -- we should look for heterotic models with
3 or 4 vectors, and try to match them to type II compactifications on known
Calabi-Yau manifolds.  After
discussing in
detail specific examples
for both the rank three and four cases,
in \S4.5 we list several more examples of
heterotic compactifications
with the right spectra to be dual to type II compactifications on known
Calabi-Yau threefolds.  We have corrected \S4.5 in light of
comments and work which appeared after the preprint version of this
paper.  Therefore, we only give examples of heterotic compactifications
with spectra which match those of type II compactifications on
Calabi-Yau threefolds which are $K3$ fibrations.  Such threefolds
appear to be the relevant class in
understanding heterotic/type II duality \KLM \VaW, and as the list of
such manifolds in \KLM\ is quite short, any matches are highly
suggestive.

\subsec{A Rank Three Example and Its Dual}

We begin our search for heterotic/type II dual pairs with a rank three
example.
Since the most familiar $K_{3}\times T^{2}$ compactifications
automatically yield
at least a rank four gauge group (from the $U(1)^{4}$ of the torus),
we must somehow remove some of the gauge symmmetries coming from
the torus to get only a rank
three gauge group at low energies.

One way of doing this is to start with the
$E_{8}\times E_{8}$ string and {\it first} compactifying to eight dimensions on
a 2-torus with $\tau=\rho$, which yields an $E_{8}\times
E_{8}\times SU(2)\times U(1)^{3}$ gauge group.
Upon further compactification on $K3$ down to four dimensions, we can
now use the extra $SU(2)$ gauge symmetry in satisfying \gencons\
by also turning on gauge fields of this $SU(2)$.
We embed $SU(2)$ bundles with
$\int_{K_3}c_{2}=10$ in
each $E_{8}$ and an $SU(2)$ bundle with $\int_{K_3} c_{2}=4$ in the
$SU(2)$.  This leaves an $E_{7}\times E_{7}\times U(1)^{3}$ gauge
symmetry.
The hypermultiplets include 3 $\bf{56}$s of each $E_{7}$ and
59 gauge neutral moduli.  Higgsing the $E_{7}$s completely yields
$2\times (3\times 56 - 133) = 70$ extra gauge neutral moduli,
leaving a spectrum of 129 hypermultiplets and 3 vectors (2 vector
multiplets).  It may appear to the reader
that we have a lot of room in choosing the $c_2$'s of various
$SU(2)$'s.  This is not so.  In fact, if we wish to break
the full $E_8\times E_8\times SU(2)$ by Higgsing there is only one
other choice!  This rigidity is in accord
with the relative scarcity of low $b_{11}$ Calabi-Yau manifolds.

We obtained this model by compactifying to eight dimensions
on a torus with $\tau=\rho$, and then breaking the resulting
$SU(2)$ enhanced gauge symmetry completely.
This removes the modulus
which would take one away from $\tau=\rho$, leaving one with a moduli
space for these $\tau=\rho$ tori which consists of only one copy of
the fundamental domain of SL(2,Z).
When $\tau=\rho=i$ there is an $SU(2)\times SU(2)$
enhanced gauge symmetry on the torus, and one of these enhanced $SU(2)$s
will still be
present in our theory.  Therefore, if we denote the
two vector multiplets in this compactification by $\tau$ and $S$ ($S$
denotes the dilaton; the third
$U(1)$ comes from the graviphoton), then we see that $\tau=i$ should be
a point where one obtains pure SU(2) N=2 gauge theory.
Thus, we are studying the closest
heterotic string analogue of the $N=2$ Yang-Mills theory recently
solved in the work of Seiberg and Witten \Seiberg.

Can we find a conjectural type II dual for this heterotic theory?
There is apparently a
unique known Calabi-Yau manifold $M$ with $b_{11}=2$ and
$b_{21}=128$.  It is the hypersurface of degree 12 in $WP^{4}_{1,1,2,2,6}$
(another realization of this manifold is discussed in \Schimm).
Given the scarcity of known Calabi-Yau manifolds with $b_{11}=2$, this
match is highly suggestive.
Luckily, the moduli space of vector
multiplets in type IIA compactification
on this manifold has been studied in great depth
in \Yau\Candelas\ using mirror symmetry.
We will now provide extremely strong evidence that the
heterotic compactification described above is dual to the type IIA string
on $M$.
In fact, we claim the moduli space of
vector multiplets on $M$ is the quantum moduli space of the dual
heterotic string!

As we remarked above, in the classical heterotic theory the moduli space
of the $\tau$ vector multiplet is given by one copy of the fundamental
domain of SL(2,Z).  In the full theory with $\tau$ and $S$ we
therefore expect that as $S \rightarrow \infty$ (weak coupling)
we should find a copy of the fundamental domain of SL(2,Z) embedded in the
moduli space, even the exact quantum moduli space.  We also expect a
singular point on this moduli space, when $\tau = i$, where classically
the $U(1)$ gauge symmetry is enhanced to $SU(2)$.

It is useful at this point to review some results of \Yau\Candelas.
The complex moduli of the mirror of $M$
can be represented roughly speaking as $\phi$ and $\psi$
in the defining polynomial
\eqn\cyhyp{p = z_{1}^{12} + z_{2}^{12} + z_{3}^{6} + z_{4}^{6} + z_{5}^{2}
- 12\psi z_{1}z_{2}z_{3}z_{4}z_{5} - 2\phi z_{1}^{6}z_{2}^{6}}
where the $z_{i}$ are the coordinates of the $WP^{4}$ (our choice of notation
in this equation for the complex moduli of the mirror follows \Candelas).
The authors of \Candelas\ actually find it convenient to introduce
the large complex structure coordinates
\eqn\largec{Y_{1} = {(12\psi)^{6}\over {2\phi}}}
\eqn\largectwo{Y_{2} = (2\phi)^{2}~.}
They then compute $Y_{1,2}$ as functions of the (exponentiated) complexified
Kahler forms $q_{1,2}$ of $M$.

We will also find it convenient to use results of \Yau, so we
need to provide a notational dictionary.
In appendix $A.1$ of \Yau, the
notation $\bar x$ and $\bar y$ is used for the
complex moduli of the mirror.  The dictionary translating between \Yau\ and
\Candelas\ is simply given by
\eqn\relyau{\bar x = -{1\over 864} {\phi\over{\psi^{6}}} = -{1728 \over Y_{1}}}
\eqn\reltwo{\bar y = {1\over \phi^{2}} = {4\over Y_{2}}~.}

We must now decide what we should check, to decide whether or not this type II
moduli
space is providing a description of the quantum moduli space of
our heterotic model.  Classically, the heterotic theory has an $SU(2)$
enhanced gauge symmetry at $\tau = i$.  Therefore we should expect to
recover a structure reminiscent of $SU(2)$ N=2 gauge theory in the weak
coupling $S\rightarrow \infty$ limit where
the gravitational effects are not significant.  More precisely, it is shown in
\Seiberg\ that in the case of $SU(2)$ N=2 gauge theory, the isolated singular
point in the perturbative theory (where $SU(2)$ is restored) splits into
two singular points in the full quantum theory, where
monopoles become massless.
So we also expect that as the string coupling is turned on, we will
see a single singular point at $\tau=i$ being split into two singular points.

Does this picture hold?  The discriminant locus of this model has been
studied by \Yau\Candelas\ where they find it is given by
\eqn\disc{(1-\bar x)^{2} - \bar x^{2}\bar y = 0~.}
The first thing we notice is that \disc\ is quadratic in $\bar x$, which is
necessary for a picture like that of \Seiberg\ to hold.  This suggests
that $\bar x$ parametrizes the $\tau$ space in some way.  At $\bar y=0$
the two solutions for $\bar x$ merge, so we should identify $\bar y = 0$ with
$S = \infty$
and $\bar x=1$ with the $SU(2)$ point.  This means that $\bar x=1$
should be identified with $\tau=i$, at least at $\bar y=0$.
Note that $\bar y=0$ corresponds to the large radius limit
of $M$, where the leading behavior of the metric on moduli space
is the same as that expected for the dilaton.

These tests have given us some idea of what to expect, in terms of
identifying the type II parametrization of the moduli space with the
heterotic parametrization.
Now we give much stronger evidence for the interpretation offered above,
based on a surprising observation of \Candelas.
There, in \S7, it was noted that at $\bar y = 0$ the mirror map giving the
relation
between $\bar x$ and the special coordinates on the
moduli space of $M$ is given by the
elliptic $j$-function!  More precisely
\eqn\almost{\bar x = {1728\over j(\tau_{1})}}
where $\tau_{1}$ parametrizes a kahler modulus of $M$ in this limit.
This was noted as an unexplained ``curiosity'' by the authors of \Candelas.
{}From \almost\ we see that, since $j(i)=1728$, the point corresponding to
$\bar x=1$ is precisely $\tau_{1}=i$!
This means we should identify
$\tau=\tau_{1}$ at least for weak coupling.
Then \almost\ implies that
\eqn\smoking{\bar x = {1728\over j(\tau)}~.}
Since $\bar y=0$ corresponds to $S = \infty$, we also propose
\eqn\idents{\bar y = e^{-S}}
to leading order in the coupling.

But we can check more.  Since we are conjecturing that the type II
vector moduli space is the fully quantum corrected version of the heterotic
vector moduli space, we should also check that the one-loop corrections
to the prepotential are reflected in the $\bar y \rightarrow 0$ limit of
the moduli space of $M$.
The one-loop correction to the prepotential $F$
in similar N=2 heterotic theories has been
computed in \Antoniadis\louis (and one could directly compare with
their computations in the next example).
For our example, predictions for the full non-perturbative third derivatives of
the prepotential directly follow from the formulas of \Yau\Candelas.
Specializing to the $\bar y \rightarrow 0$ limit, we should be able to
recover the string one-loop corrections.
In particular, using the formulas of appendix $A.1$ of \Yau\ (and remembering
to restore the factor of the discriminant
\disc\ in the denominators!) and the dictionary \smoking \idents , we see we
are
predicting
\eqn\tttyuk{F_{\tau\tau\tau} = {j_{\tau}^{3} \over {j(\tau)
(j(\tau)-j(i))^{2}}}}
\eqn\ttsyuk{F_{S\tau\tau} = {j_{\tau}^{2} \over {j(\tau)(j(\tau)-j(i))}}~}
where $j_\tau$ denotes the derivative of $j$ with respect to $\tau$.
$F_{SS\tau}$ and $F_{SSS}$ vanish in the limit of weak coupling that we
are considering, as expected from perturbative string theory.

Actually, the formulas \tttyuk\ and \ttsyuk\ are not quite the final story.
In string theory, we expect the gauge coupling function, obtained by
taking two derivatives of the prepotential, to really be a {\it function}
(and
not a section of some bundle).
This means that the third derivatives $\tilde F_{\tau\tau\tau}$ and
$\tilde F_{S\tau\tau}$ for
our example should really be a modular
forms of weight 2 and 0 with
respect to the SL(2,Z) symmetry acting on $\tau$.
This has been obscured by the gauge choice made for
the `Yukawa couplings',
but we can restore the correct modular properties by simply recognizing
that the appropriate gauge for the `Yukawa couplings'  must be\foot{
It would be desirable to see why this is natural also from the type II side
using the results of \Yau \Candelas .  Note that
the gauge independent quantity $F_{\tau\tau \tau}/
F_{S\tau \tau}=j_\tau/(j(\tau) -j(i))$ agrees with what
one expects from the heterotic string.}
\eqn\trueprep{\tilde F_{\tau\tau\tau} = {F_{\tau\tau\tau}\over E_{4}(\tau)} }
\eqn\truepreps{\tilde F_{S\tau\tau} = {F_{S\tau\tau} \over E_{4}(\tau)}}
where $E_{4}(\tau)$ is the fourth Eisenstein series
(which is
$1/240$-th of the theta function for the $E_{8}$ lattice), and we know
it must appear due to the uniqueness of weight four modular forms.
So the true formulas we are predicting for the third derivatives of the
prepotential are:
\eqn\trueyukt{\tilde F_{\tau\tau\tau} = {j_{\tau}^{3}\over {E_{4}(\tau)
j(\tau)(j(\tau)-j(i))^{2}} }}
\eqn\trueyuks{\tilde F_{S\tau\tau} = {j_{\tau}^{2}\over {E_{4}(\tau)j(\tau)
(j(\tau)-j(i))}}~.}

How do \trueyukt\ and \trueyuks\ compare with expectations?
Based on its singularity properties and asymptotic behavior alone (for similar
arguments see e.g. \kap\Antoniadis\louis), we
know we can fix the
gauge coupling function $\tilde F_{\tau\tau}$ to be
\eqn\gcf{\tilde F_{\tau\tau} \sim S+log (j(\tau) - j(i))~.}
This means that we expect
\eqn\expec{\tilde F_{\tau\tau\tau} \sim {j_{\tau} \over {(j(\tau)-j(i))}}~.}
Remarkably, the elliptic $j$-function satisfies the identity
\eqn\identity{j_{\tau}^{2} = -960 \pi^2 j(\tau)(j(\tau)-j(i))E_{4}(\tau)}
which makes it clear that \trueyukt\ is in fact of the expected form
\expec !
Similarly, we see using the identity \identity\ that $F_{S\tau\tau} \sim 1$
at weak coupling, as expected.
So we see our results agree not only with the classical heterotic string
picture but also with the expected one-loop string correction.

In identifying the algebraic variables $\bar x$ and $\bar y$ with
$\tau$ and $S$ beyond the leading order it is natural
to conjecture that they continue to be related by the mirror map.

\subsec{New Stringy Phenomena}

So far we have given strong evidence for the identification of our heterotic
model with the Type IIA string on $M$.
Our main interest is to use these results to
see how nonperturbative string corrections modify the classical string
picture.  For example, there might be
qualitatively new effects due to the presence of
gravity.  Here we will make a few preliminary remarks, leaving the full
story to future work.

For finite $\bar y$, the singular locus already ``knows'' that the moduli
space is the fundamental domain of $SL(2,Z)$.   But for very small $\bar y$
this may not be the case.
In this region of infinitesimal $\bar y$, we can hope to
recover in much greater detail the results of \Seiberg (perhaps
by considering a double scaling limit).
We might also see stringy corrections to these results.
This is presently under investigation \ref\KV{S. Hosono, S. Kachru and
C. Vafa, work in progress.}.

Alternatively, for
finite $\bar y$ we might also probe qualitatively new stringy
modifications due to nonperturbative effects present in string theory
but absent in field theory.  In fact, in addition to the discriminant
locus \disc\ which is the locus of conifold singularities, there is an
additional singular locus at $\bar y=1$ where the manifold acquires
a complicated point singularity \Candelas.\foot{We would like to thank
S. Hosono for helpful correspondence on this point.}  Note that with the
identification \idents\  the locus $\bar y =  1$ corresponds to
$S\rightarrow 0$, i.e. infinitely strong coupling.
Moreover, along this locus
the threefold becomes birationally equivalent to the $(2,6)$ complete
intersection in $WP^{5}_{1,1,1,1,1,3}$ \Candelas , suggesting that there should
be a smooth transition similar to \stromtwo.  It would be interesting
to unravel the physics of this transition, from the type II side.
One concrete hint is the singularity structure in the
one-loop computation of $R^{2}$, which gives the
net number of massless hypermultiplets \Vafatest.

An interesting point to notice is that there is an extra $Z_{6}$
symmetry for all $\bar y$ at $\bar x = \infty$, which at $\bar y=0$
corresponds to $\tau = {{1\over 2} + i{{\sqrt 3} \over 2}}$.  This
is the quantum $Z_{6}$ symmetry of the Landau-Ginzburg theory \Candelas, and
we see that it survives string nonperturbative effects.
Moreover, if we tune the coupling constant properly (and with a particular
choice of $\tau$ for the heterotic compactification) we get an enhanced
$Z_{12}$ symmetry point on moduli space (corresponding to
$\phi=\psi=0$).

\subsec{A Rank Four Example and its Dual}

Having met with success in finding a dual for a rank three example, we
now move on to a rank four example.
Let us start with the heterotic
$E_{8}\times E_{8}$ string.  Embed a rank 2 bundle with
$\int_{K_{3}}c_{2}(V)=12$ in each $E_{8}$; this gives rise to a theory with
4 $\bf{ 56}$s in each $E_{7}$ and a total of 62 gauge neutral moduli
hypermultiplets.  Now Higgs both $E_{7}$s completely by giving VEVs
to the charged fields.  This gives an extra $4\times 56 - 133 = 91$
neutral fields from each factor, leaving us with 244 hypermultiplets and
just the $U(1)^{4}$ gauge symmetry of the torus (a $U(1)^{2}$ of which
is enhanced to
$SU(2)\times U(1)$, $SU(2)\times SU(2)$ and $SU(3)$ at special points).

We would like to find a Calabi-Yau manifold $X$ with $b_{11}=3$ and
$b_{21}=243$, on which type II strings could be dual to the heterotic
theories we have just described.  And in fact, such a manifold $X$ does
exist!
It is the degree 24 hypersurface in $WP^{4}_{1,1,2,8,12}$ which has been
studied (using mirror symmetry) in \Yau.

$X$ is defined by an equation of the form
\eqn\yaueq{a_{1}z_{1}^{2} + a_{2}z_{2}^{3}+a_{3}z_{3}^{12}+a_{4}z_{4}^{24}+
a_{5}z_{5}^{24} - 12\alpha z_{1}z_{2}z_{3}z_{4}z_{5} - 2\beta z_{3}^{6}
z_{4}^{6}z_{5}^{6} - \gamma z_{4}^{12}z_{5}^{12}}
where the $z_{i}$ are the weighted projective space coordinates.
It is actually convenient to define
\eqn\trueparx{\bar x = -{1\over 3456}{a_{1}^{3}a_{2}^{2}\beta \over
{\alpha^{6}}}}
\eqn\truepary{\bar y = {4a_{4}a_{5}\over {\gamma^{2}}}}
\eqn\trueparz{\bar z = -{1\over 2}{a_{3}\gamma\over {\beta^{2}}}}
which serve as the complex structure coordinates of the mirror.

There are several things we expect to be true for this model, which we
would like to check.  For example, we expect that in an
appropriate weak coupling
limit, there should be a copy of the $\tau$ and $\rho$ moduli spaces
of $T^{2}$ (a product of two
copies of the fundamental domain of $SL(2,Z)$) embedded
in the Kahler moduli space of $X$.
In this weak coupling limit, we can also make several statements about
the singularity structure.
At generic points in the moduli space, the left-movers of the toroidal
compactification are responsible for a $U(1)^{2}$ gauge symmetry.
But on the locus $\tau=\rho$ this is enhanced to an
$SU(2)\times U(1)$ gauge symmetry while the points $\tau=\rho=i$
and $\tau=\rho={{1\over 2} + i{{\sqrt 3} \over 2}}$ have further
enhancement to $SU(2)\times SU(2)$ and $SU(3)$
gauge symmetry, respectively.
We therefore expect a singular locus, at very weak coupling, which looks like
the $\tau=\rho$ copy of the fundamental domain of $SL(2,Z)$, with two
special points.

We now give a description of the good coordinates on the moduli space.
The SL(2,Z) invariances tell us that we should work with $j(\tau)$ and
$j(\rho)$, as they map the fundamental domain of SL(2,Z) bijectively
to the complex plane.
There is also a $Z_{2}$
exchanging $\tau$ and $\rho$, so the really good coordinates are
\eqn\udef{u = j(\rho) + j(\tau)}
\eqn\vdef{v = 4j(\rho)\times j(\tau)~.}
The codimension one singularity, at weak coupling, should be at
$\tau=\rho$ which is the same as the locus
\eqn\discpred{u^{2} - v = 0~.}
We expect to have a double point at weak coupling when $\tau=\rho$ for
$\it generic$ $\tau$ and $\rho$, since we have an $SU(2)$ enhanced
gauge symmetry.  So we really expect a discriminant locus of the form
\eqn\discpredt{(u^{2} - v)^{2} = 0}
at zero coupling.

The discriminant locus of $X$ is given by \foot{There is a misprint
in the discriminant locus presented in appendix $A.1$ of the
hep-th version of \Yau, corrected in the published version.  We thank
S. Hosono for informing us of this.}
\eqn\disctwo{(1-\bar x)^{4} - 2\bar x^{2} \bar z (1-\bar x)^{2}+
\bar x^{4} \bar z^{2} (1-\bar y)=0~. }
First of all, note that at $\bar y=0$ the equation becomes
\eqn\weakdisc{((1-\bar x)^{2} - \bar x^2 \bar z)^{2} = 0}
which is of the form \discpredt\ expected from the heterotic string, where
roughly speaking $u \sim \bar x$ and $v \sim \bar z$.  To check in more
detail this duality, one would like to find a relationship at
$\bar y = 0$ between $\bar x$, $\bar z$,
and the $j$ functions.
At our request, Hosono has investigated this question using the results of
\Yau\ and has found some preliminary evidence suggesting such a relationship.
Very detailed checks for this example must await further investigation of
this question \KV.
Nevertheless, we will try to proceed with some further qualitative checks.

Another qualitative
fact we would like to
explain is the presence of codimension two singularities which should
occur (at weak coupling) at the $SU(2)\times SU(2)$ and $SU(3)$ points.
In fact, there are further singular loci in the moduli space of $X$,
one of which is given
by
\eqn\loctwo{ (1-\bar z)^{2} - \bar y \bar z^{2} = 0~.}
This intersects weak coupling ($\bar y = 0$) at $\bar z = 1$.
Furthermore, there is an intersection of this locus with the
$\tau=\rho$
locus at two points.  One intersection
occurs at $\bar x = \infty, \bar z = 1$ where
there is a $Z_{6}$ symmetry; we would like to identify this with the
$SU(3)$ point.  The other intersection is at $\bar x = {1\over 2},
\bar z = 1$.  We would like to identify this point with the $SU(2)\times SU(2)$
point.  The symmetry properties of this moduli space unfortunately have
not been worked out.  This would be crucial for a better understanding
of the nature of the $\bar z=1$ singularity.  However, for generic
$\tau$, $\rho$ with $\epsilon = \tau - \rho$ small, we have checked
using the results of \Yau\
that the requisite singularity structure of the third derivatives of
the prepotential is reproduced.  In particular,
\eqn\checkone{F_{\epsilon\epsilon\epsilon} \sim {1\over \epsilon}}
\eqn\checktwo{F_{\epsilon\epsilon S} \sim 1}
with $F_{\epsilon SS}$ and $F_{SSS}$ vanishing.

\subsec{More Potential Dual Pairs}

Since we have been studying very special examples, the reader might wonder
how easy it is to find other examples where one can propose a (testable)
duality between heterotic and type II compactifications.
As mentioned in \S4.1, this subsection has been revised since the
preprint version of this paper.  In light of the
very recent work
suggesting that CY threefolds which are K3 fibrations play an important
role in heterotic/type II duality,
it makes sense to try and find heterotic duals for the K3 fibrations
listed in \KLM\ (which is a highly restricted list, compared to the
lists of all known CY spaces).
At the level of
simply matching the numbers of hypermultiplets and vector multiplets,
it is not difficult to construct many such examples, and we
provide some below (all on the list in \KLM). It
would be
extremely interesting to explore
some of these examples in more detail to see
if one's expectations for the heterotic models are reflected in the type II
moduli spaces, as in \S4.2.
Of course our construction of examples is by no means
exhaustive.  It might prove beneficial
to check other sorts of examples, or to find a general recipe linking
the heterotic constructions to the dual manifolds.

One nice class of heterotic N=2 compactifications
can be constructed as follows.   Start with a compactification to
nine dimensions on a Narain lattice $\Gamma^{1,17}$ which gives
$SO(34)\times U(1)$ gauge group.  Further compactify to eight dimensions
on a circle, including Wilson lines which break the $SO(34)$ to
an $SO(34-2n)\times SO(2n)$ in a way consistent with level-matching \level.
This yields a theory in eight dimensions with $SO(34-2n)\times SO(2n)\times
U(1)^{3}$ gauge group, and upon compactification to four dimensions
on $K_{3}$ we can compute the resulting spectrum using the techniques of
\S2.  We call the factors of the vacuum gauge bundle embedded in
$SO(34-2n)$ and $SO(2n)$ $V_{1}$ and $V_{2}$ (where we
keep the convention $n\leq 8$), and
we use the notation
\eqn\notation{ d_{1} = \int_{K_{3}}c_{2}(V_{1}),~~d_{2} = \int_{K_{3}}
c_{2}(V_{2})~.}

Some examples which fall into this category are the following:
\medskip

\noindent 1)  Consider the $n=6$ case.
In the further compactification
on $K_{3}$, choose $V_{1,2}$ to be
$SU(2)$ bundles with $d_{1,2}$ = $16$ and $8$.
After Higgsing
maximally, leaving an $SO(6)\times SO(4)\times U(1)^{3}$ gauge
group unbroken, the spectrum at generic points in the moduli
space is $(144,8)$.
There is a known K3 fibration with $b_{11}=7, b_{21}=143$ --
the hypersurface in $WP^{4}_{1,1,4,4,10}$.
\medskip
\noindent 2)  Consider the $n=6$ case, and
in the further
compactification on $K_{3}$ choose $V_{1}$ to be a $d_{1}=20$
$SU(2)$ bundle
and $V_{2}$ to be a $d_{2}=4$ $SU(2)$ bundle.
After Higgsing as much as possible one is left with
a generic spectrum of $(195,9)$.  The Calabi-Yau hypersurface in
$WP^{4}_{1,1,4,8,14}$ is a K3 fibration with $b_{11}=8, b_{21}=194$.
\medskip
\noindent 3)  Consider the construction above, with the compactification
to 8 dimensions done on a circle at the self-dual radius with no
Wilson lines.  Then the gauge group is $SO(34)\times SU(2)\times
U(1)^{2}$.
Now in compactifying on K3 to four dimensions, embed rank two
bundles with $\int_{K3} c_{2} = 20$ and $4$ into the $SO(34)$
and $SU(2)$, respectively.  Maximally Higgsing yields a spectrum
of $(195,9)$ again, so as in example 2) this model is appropriate for a dual
to the hypersurface in $WP^{4}_{1,1,4,8,14}$.  In particular, it should
lie in the same moduli space as example 2).
\medskip
We can give more examples by following the same strategy we used
with $SO(34)$, this time with $SO(32-2n)\times SO(2n) \times U(1)^{4}$.  We
adhere to the same notation,
and find the following examples which have known candidate Calabi-Yau
duals:
\medskip
\noindent 4) Take $n=0$, compactifying on $K_{3}\times T_{2}$ in
the normal way.  Take $V_{1}$ to be a rank 2 bundle with $d_{1}=24$.
One is left with $SO(28)\times SU(2)\times U(1)^{4}$ gauge group
in four dimensions, with 10 ${\bf {(28,2)}}$s of $SO(28)\times SU(2)$.
Higgsing as much as possible, one is left with an unbroken $SO(8)\times
U(1)^{4}$ gauge group and generically a
$(272,8)$ spectrum.  This is appropriate
for
a dual to the type IIA string compactified on the hypersurface
in $WP^{4}_{1,1,4,12,18}$.  This example (and the following one)
was mentioned in \VaW, where an error in the preprint
version of this paper was corrected.
\medskip
\noindent 5) Take
$n=0$, that is consider the $SO(32)$ string
compactified to eight dimensions on a torus, but take the torus to
sit at $\tau=\rho={1\over 2} + i {{\sqrt 3}\over 2}$.  Then
the gauge symmetry is $SO(32)\times SU(3)\times U(1)^{2}$.
Embed an $SU(2)$ bundle with $\int_{K_{3}}c_{2}=18$ in the $SO(32)$
and an $SU(3)$ bundle with $\int_{K_{3}}c_{2}=6$ in the $SU(3)$.
After Higgsing, one obtains the spectrum $(165,9)$ which would
arise from type IIA compactification on a manifold with $b_{11}=8,
b_{21}=164$.  Such a K3 fibration does exist -- the hypersurface
in $WP^{4}_{1,1,4,6,12}$.
\medskip
We can also use the $E_{8}\times E_{8}$ string as well
as the more general groups one can get in Narain compactification
to eight dimensions to try and generate more
examples.
Here we present some
simple examples using these other possibilities.  Below,
$V_{1,2}$ and $d_{1,2}$
refer to the bundles embedded in the two $E_{8}$s.
\medskip
\noindent 6) Start with the ``standard embedding'' compactification on
$K3\times T^{2}$, i.e. with $V_{1}$ an $SU(2)$ bundle and $d_{1}=24$.
Now Higgs the unbroken $E_{7}$ completely --
the resulting spectrum $(492,12)$ would arise from type IIA strings on
a Calabi-Yau with $b_{11}=11, b_{21}=491$ and there is such a space,
the hypersurface in $WP^{4}_{1,1,12,28,42}$ (note that $b_{21}$ of this space
is given incorrectly in the list of \KLM, although $b_{11}$ and the
Euler character are given correctly there).
\medskip
\noindent 7) Start with $E_{8}\times E_{8}\times SU(2)\times U(1)^{3}$
in eight dimensions by compactifying on a torus with $\tau=\rho$.
In the further reduction on K3 choose $V_{1}$ a rank two bundle
with $d_{1}
= 20$ and also embed a rank two bundle
with $\int_{K3} c_{2} = 4$ into
the $SU(2)$.   After Higgsing here, one obtains a $(377,11)$
spectrum appropriate to a potential dual for the hypersurface
in $WP^{4}_{1,1,8,20,30}$.
\medskip
\noindent 8) Start with $E_{8}\times E_{8}\times SU(3)\times U(1)^{2}$
by compactifying to eight dimensions on a torus with
$\tau=\rho = {1\over 2} + i {{\sqrt 3}\over 2}$.
Choose $V_{1}$ and $SU(2)$ bundle with $d_{1} = 18$
and also embed rank three bundle with $\int_{K3} c_{2} =
6$ in the $SU(3)$ when you further compactify on K3.
After Higgsing this is a potential dual to the $b_{11}=9, b_{21}=321$
manifold on the list in \KLM, the hypersurface in $WP^{4}_{1,1,6,16,24}$.
\medskip
\noindent 9) Start in eight dimensions just as in example 8), but
this time choose $V_{1,2}$ both of rank two with $d_{1} = 10, d_{2}=8$,
and also
embed an $SU(3)$ bundle with
$\int_{K3}c_{2}=6$ into the $SU(3)$.  Completely Higgs the
first $E_{8}$ and Higgs the other down to $SO(10)$, then go
to a generic point in the moduli space of the vectors.  This
yields a potential dual to the complete intersection of
degree 8 and 12 hypersurfaces in $WP^{5}_{1,1,2,4,6,6}$,
which has $b_{11}=6, b_{21}=98$.
\medskip
\noindent 10) Start with the Narain compactification yielding
$E_{8}\times SO(20)\times U(1)^{2}$ gauge group in eight dimensions.
Embed rank two bundles with $\int_{K3} c_{2} = 10$ and $14$
into the $E_{8}$ and the $SO(20)$ factors in the further
compactification on K3.  Higgsing as much as possible
(leaving an $SO(6)$ subgroup of the $SO(20)$ unbroken), one
finds a generic spectrum $(149,5)$, which could describe
a heterotic dual to type IIA strings on the hypersurface in
$WP^{4}_{1,1,2,4,8}$.
\medskip
\noindent 11)  Start with the compactification on a special torus yielding
$E_{8}\times E_{8}\times SU(3)\times U(1)^{2}$ gauge
group in eight dimensions.
Choose $V_{1,2}$ to be rank two bundles with $d_{1} = 10$ and $d_{2} = 8$,
and embed a rank three bundle
with $\int_{K3}c_{2} = 6$ into the $SU(3)$.  After Higgsing
the first $E_{7}$ completely and the second $E_{7}$ down to $SO(8)$, move
to a generic point in the moduli of the vectors.
This leaves a spectrum of $(102,6)$, appropriate
for the dual of the type IIA string compactified on the Calabi-Yau
hypersurface in $WP^{4}_{1,1,2,4,4}$ which has $b_{11}=5, b_{21}=101$.
\medskip
Looking through the list of threefold K3 fibrations for which we have found
potential heterotic duals, one is struck by certain phenomenological
patterns which the weights of the relevant weighted projective spaces
exhibit.
For example, starting from some examples with low $b_{11}$ and shifting
the weights of the ambient $WP^{4}$ by $(0,0,2,4,6)$ often yields another
example.
Hopefully this is an indication that a general recipe
connecting certain classes of Calabi-Yaus to their heterotic duals
is within our reach in the near future.

We would like to close by discussing one other type II
model which we believe may
be describing the nonperturbative structure of $SU(2)$ N=2 gauge theory
with $N_{f}=1$ matter hypermultiplet in the $\bf {2}$ of $SU(2)$.
In such a theory, we expect that the classical moduli space of the vectors
will exhibit a single singular point (where $SU(2)$ is restored and the
charged matter is massless), which splits into $\it three$ singular
points in the quantum theory \SWTwo.
The Kahler moduli space of the hypersurface in $WP^{4}_{1,1,1,6,9}$,
studied in depth in \Yau\Katz, is a perfect candidate for the
string description of this moduli space.  In the coordinates of \Yau,
its discriminant locus is given by
\eqn\mattdisc{(1-\bar x)^{3} - \bar x^{3} \bar y = 0~.}
Identifying $\bar y$ with the dilaton in the usual way \idents, we
see that this is precisely the singularity structure we expect for N=2
$SU(2)$ gauge theory coupled to a single $\bf 2$ of $SU(2)$.
It would be very interesting to find the heterotic dual of the type II
compactification on this Calabi-Yau hypersurface.

\newsec{Conclusions}

We have seen, through examples, that it is possible that the moduli spaces
of many different N=2 heterotic vacua are connected in a large web, in a
similar way to type II string compactifications \stromtwo.   For some
of these N=2 heterotic models, we have also been able to construct dual
type II string compactifications on Calabi-Yau threefolds.

This raises many interesting questions.  The first thing one might wonder about
is the generality of the phenomenon -- when do we expect a compactification
to have dual heterotic and type II descriptions?  Given the bound on the rank
of the gauge group attainable in perturbative heterotic strings, it seems
unlikely that type II compactifications on Calabi-Yau manifolds with
sufficiently large hodge numbers will have heterotic duals.  Similarly,
we have found examples of heterotic compactifications
for which there are no presently known candidate Calabi-Yau
duals.
In fact, the $(0,24)$ example of \S3 is an example for which there
$\it cannot$ be a dual Calabi-Yau description -- that would require
$b_{12} = -1$!  In other words, such a theory has no dilaton on the
type II side. What we
envision is that the whole moduli space of N=2, d=4 string theories
form a connected web.  Some regions have type II descriptions, some regions
have heterotic descriptions, some regions have both (as we have found),
and perhaps some regions have neither.  Consider for example the
$E_{8}\times E_{8}$ string
broken to $E_{7}\times E_{8}$ with the standard embedding,
as in example 6) of \S4.5.  At a Gepner point in
its moduli space with enhanced $U(1)^{5}$ gauge symmetry, we can move
(as discussed in \S3)
off onto a new branch of moduli space which gives rise to the spectrum
$(0,24)$.  On the other hand, we saw that by Higgsing we could get to
a theory with spectrum $(492,12)$ for which we have proposed a dual type II
Calabi-Yau compactification with $b_{11}=11, b_{21}=491$!
This means that there is a path one can follow from the $(492,12)$ type II
theory which has a Calabi-Yau description to a region with spectrum
$(0,24)$ which cannot have a Calabi-Yau description.  Similarly,
starting from the Calabi-Yau with $b_{11}=11, b_{21}=491$ we can
move using conifold transitions \stromtwo\confo\ to a Calabi-Yau with
both $b_{11}+1$ and $b_{21}+1$ larger than the allowed rank 24 of a gauge group
in perturbative heterotic strings.   We could follow the same path
starting from the $(492,12)$ heterotic string, reaching a theory without
a perturbative heterotic string description!

In this paper, we have focused on using the duality between certain
heterotic and type II N=2 compactifications to determine the exact
structure of the moduli space of vectors on the heterotic side in terms
of the tree level structure on the type II side.  Similarly, one could
compute the exact structure of the moduli space of hypermultiplets on
the type II side using just the tree level results for the moduli space
of the heterotic hypermultiplets.
Completely understanding the
tree level structure on the heterotic side would involve
answering certain
questions about the moduli spaces of stable
holomorphic vector bundles over $K_{3}$ and about the moduli spaces of
Higgs fields (which are probably related as we can describe
the same moduli space in two different ways).

For examples where we have been successful in constructing a dual, there are
many things one would like to learn.  There is the exciting possibility of
probing stringy non-perturbative effects in a very quantitative manner,
and perhaps uncovering qualitatively new physics.
Perhaps such effects would have counterparts in other compactifications,
for example compactifications with only $N=1$ supersymmetry.

\centerline{\bf {Acknowledgements}}

We would like to thank S. Hosono for
extremely helpful correspondence.  We would also like to thank
S. Ferrara, J. Harvey and A. Strominger for sharing their insights on
this problem.
We have also benefited from interesting discussions with N. Berkovits, A.
Klemm, R. Plesser, H. Sarmadi, S. Shenker,
and S.T. Yau.
The work of SK is supported in part by the
Harvard Society of Fellows and the William F. Milton Fund of Harvard
University. The work of CV is supported in part by NSF grant
PHY-92-18167.

\listrefs
\end